\documentclass{article}
\usepackage{graphicx}  
\usepackage{amsmath}   
\usepackage[compress,numbers,sort]{natbib}
\usepackage{amssymb}   
\usepackage{bm} 
\usepackage{dcolumn}
\usepackage{color}
\usepackage{mathrsfs}
\usepackage{amsfonts}
\usepackage{varioref}
\usepackage{textcomp}
\usepackage[normalem]{ulem}

\usepackage{multirow}
\usepackage{caption}
\usepackage{subcaption}
\RequirePackage[colorlinks,citecolor=blue,urlcolor=magenta,linkcolor=blue]{hyperref}
\allowdisplaybreaks
\labelformat{section}{Section #1} 
\labelformat{subsection}{Section #1} 
\labelformat{subsubsection}{Section #1}
\labelformat{subsubsubsection}{Section #1}
\labelformat{equation}{Eq.~(#1)} 
\labelformat{figure}{Fig.~#1} 
\labelformat{subfigure}{Fig.~\thefigure#1} 
\labelformat{table}{Table~#1} 
\labelformat{appendix}{Appendix #1}

\title{\bf On the stability of the objects of limiting compactness: Black hole and Buchdahl star }

\author{ Soumya Chakrabarti \footnote{soumya.chakrabarti@vit.ac.in}$~^{1}$, Chiranjeeb Singha\footnote{chiranjeeb.singha@iucaa.in}$~^{2}$ and Naresh Dadhich\footnote{Prof. Naresh Dadhich passed away on the 6th of November, 2025. The whole idea and calculations were thoroughly discussed among all the authors before his death, except for the final drafting of the manuscript.}$~^{2,\;3}$\\
$^{1}$ \small{School of Advanced Sciences, Vellore Institute of Technology, Tiruvalam Rd, Katpadi,}\\ \small{ 
 Vellore, Tamil Nadu 632014, India}\\
$^{2}$\small{Inter-University Centre for Astronomy \& Astrophysics, Post Bag 4, Pune 411 007, India}\\
$^{3}$\small{Astrophysics Research Centre, School of Mathematics,
Statistics and Computer Science,}\\
\small{ University of KwaZulu-Natal,
Private Bag X54001, Durban 4000, South Africa}
}

\begin{document}
  
\maketitle
\begin{abstract} 
In General Relativity, there exist two objects of limiting compactness, one with a null boundary defining the horizon of a black hole and the other with a timelike boundary defining a Buchdahl star. The two are characterized by gravitational energy equal to or half the mass. Since non-gravitational mass-energy is the source of gravitational energy, both of these objects are manifestly stable. We demonstrate in this letter, in a simple and general way, that the equilibrium state defining the object is indeed stable, independent of the nature of the perturbation. 
\end{abstract}

\section{Introduction}

The characterization of equilibrium and stability in highly compact relativistic objects is a long-standing and fundamental problem in gravitational physics. In regions of extreme densities, relativistic effects dominate and impose universal constraints that are largely independent of the microscopic properties of matter. In four-dimensional General Relativity (GR), a gravitational collapse is the only known process through which a regular distribution of matter evolves towards extreme density, and black holes represent the expected endpoint of a continued collapse \cite{OppenheimerSnyder1939, Israel1967, Carter1971}. Regular stellar configurations are subject to an upper bound on compactness, known as the Buchdahl limit \cite{Buchdahl1959, Bondi1964}. Although these systems are physically distinct, one characterized by the presence of an event horizon and the other supported against collapse by material stresses, both saturate fundamental bounds imposed by the Einstein equations. This suggests that their equilibrium and stability properties may be governed by deeper geometric or energetic principles common to strongly gravitating systems \cite{Wald1984, Padmanabhan2010}.  \\

Understanding these principles is particularly important given the wide range of compact objects encountered in relativistic astrophysics. While black holes are uniquely determined by a small number of global charges \cite{Israel1967,Carter1971}, regular compact stars exhibit a rich dependence on internal structure and equation of state. Nevertheless, the existence of universal compactness bounds and limiting configurations indicates that stability at extreme gravity may admit a description that transcends detailed matter models \cite{Chandrasekhar1964,Weinberg}. Identifying a unified framework capable of describing both black holes and maximally compact stars, therefore, remains an important open problem. Conventional analyses of the stability of compact objects rely primarily on linear perturbation theory, often formulated in terms of radial oscillations or normal mode analysis \cite{Chandrasekhar1964, ShapiroTeukolsky, ThorneCampolattaro1967}. While these approaches provide valuable dynamical insight, they depend sensitively on the assumed equation of state and detailed properties of matter fields. Moreover, perturbative methods are inherently local and dynamical, which can obscure the global geometric and energetic mechanisms underlying equilibrium \cite{FriedmanStergioulas}. These limitations motivate the search for a non-perturbative, equation-of-state-independent criterion for stability, rooted directly in the structure of spacetime itself.  \\

In this context, quasi-local notions of gravitational energy provide a natural and physically meaningful framework. Unlike global quantities such as ADM mass \cite{ArnowittDeserMisner}, quasi-local energies characterize the energetic content of finite spacetime regions and incorporate both matter and gravitational contributions in a unified manner \cite{Szabados2009}. Among the various prescriptions proposed in the literature, the Brown-York quasi-local energy \cite{Brown:1992br} is particularly well-suited for the analysis of compact gravitating systems. Defined within the Hamiltonian formulation of GR, it yields a covariant and finite energy associated with a closed two-surface and is directly related to the boundary terms in a gravitational action \cite{York1986}. Quasi-local energy has been successfully explored in literature, in the context of black hole thermodynamics, gravitational collapse, and dynamical spacetimes \cite{BoothMann1999, Hayward1994}. \\

In this Letter, we demonstrate that Black holes and Buchdahl stars, the two configurations of limiting compactness in GR, can be unified within a single description based on the Brown-York quasi-local energy. By identifying the gravitational field energy as the difference between quasi-local energy and non-gravitational mass-energy enclosed within a given radius, we show that both systems satisfy a universal balance condition,
\begin{equation}
E_{\rm GF} = k\, E_M ,
\end{equation}
with $k=1$ for black holes and $k=\tfrac12$ for Buchdahl stars \cite{Dad22, Dadhich:2023csk, Chak22, Shaymatov:2022ako, Shaymatov:2022hvs, Mak:2001eb, article, Andreasson:2012dj, Stuchlik:2000gey, Karageorgis:2007cy, Andreasson:2007pj, Germani:2001du, Garcia-Aspeitia:2014pna, Goswami:2015dma, PoncedeLeon:2000pj, Wright:2015yda, 2009GReGr..41..453Z,  Dadhich:2016fku, Feng:2018jrh}. Remarkably, this relation depends only on the global energetic structure of the spacetime and is entirely independent of the microscopic properties of matter, reflecting the universal character of gravitational self-energy \cite{MisnerSharp1964, Padmanabhan2010}. Our central result is that the condition of limiting compactness corresponds to an extremum, specifically, a minimum, of an appropriate quasi-local energy functional. This establishes a non-perturbative and equation-of-state-independent criterion for stability, closely analogous to variational principles encountered elsewhere in gravitational physics \cite{ReggeTeitelboim, Heusler1996}. In this framework, equilibrium and stability emerge as energetic necessities dictated by the gravitational field equations, rather than as consequences of specific perturbative dynamics. The criterion can be applicable not only to static configurations but to charged, rotating, and higher-curvature theories of gravity, including Lovelock generalizations \cite{Lovelock1971, Padmanabhan2013} as well. We argue that the extremization principle provides a natural bridge between static equilibrium and a slow self-gravitating evolution. In the quasi-static regime, relevant to the early stages of gravitational collapse or adiabatic accretion, each constant-time slice of an evolving spacetime can be regarded as an instantaneous equilibrium configuration, with stability governed by the same quasi-local energy extrema \cite{BoothFairhurst2007, AshtekarKrishnan2004}. This perspective clarifies why objects saturating limiting compactness arise as dynamically selected and energetically stable endpoints of a collapse. Overall, our results highlight the utility of quasi-local gravitational energy as a fundamental diagnostic for compactness, equilibrium and stability in relativistic self-gravitating systems \cite{Szabados2009, Padmanabhan2010}.\\

The paper is organized as follows. In \ref{stability}, we analyze the stability of a static compact object. In \ref{connection}, we establish the connection between stability and the extremization of canonical energy. The stability analysis for non-static compact objects is presented in \ref{sec:BY_conformal}. Finally, in \ref{con}, we summarize our results.  \\

\textit{Notations and conventions:} Throughout this work, we adopt the mostly plus metric signature. Accordingly, in $1+3$ dimensions, the Minkowski metric in Cartesian coordinates is $\mathrm{diag}(-1,+1,+1,+1)$. Unless otherwise stated, all calculations are performed in geometrized units, as defined earlier.

\section{Stability for a Static Compact Object} \label{stability}
In the Brown-York approach, a spacetime region is enclosed by a 3-dimensional, timelike cylindrical surface, which is itself bounded by 2-dimensional surfaces at each end. The Brown-York quasi-local energy is then defined as \cite{Brown:1992br},
\begin{equation}\label{BY_GR_Eq01}
E_{\rm BY}= \frac{1}{8\pi} \int{d^2x\sqrt{q}(k-k_0)}~,
\end{equation}
where $k$, $q_{ab}$ are the trace of extrinsic curvature and metric on $2$-surface respectively. $k_0$ is used to denote a reference spacetime. We derive the Brown-York energy of a star in GR. The exterior spacetime of a compact object can be described by the Schwarzschild geometry, the unique spherically symmetric vacuum solution of Einstein's field equations. The line element is given by
\begin{equation}\label{Schwarzschild1}
ds^{2}=-f(r) dt^{2}+\frac{d r^2}{f(r)}+ r^2(d \theta^{2}+\sin^2\theta)~.
\end{equation}
Here $f(r)$ is given by,
\begin{eqnarray}
    f &=& \bigg(1-\frac{2M}{r}\bigg)~.
\end{eqnarray}
This metric is asymptotically flat, with $f(r) \to 1$ as $r \to \infty$. We define the boundary hypersurface $\Sigma$ by a $t = \textrm{constant}$ hypersurface and the two-surface $\mathcal{B}$ as a $r = \textrm{constant}$ hypersurface within $\Sigma$. Then the unit timelike normal $u_{a}$ to $\Sigma$ and unit spacelike normal $n_{a}$ to $B$ turn out to be 
\begin{align}
u_{a}&=-\sqrt{f}\left(1,0,0,0\right);\qquad u^{a}=\frac{1}{\sqrt{f}}\left(1,0,0,0\right)
\label{BY_GR_Eq05}
\\
n_{a}&=\frac{1}{\sqrt{f}}\left(0,1,0,0\right);\qquad n^{a}=\sqrt{f}\left(0,1,0,0\right)~.
\label{BY_GR_Eq04}
\end{align}
Using the normal to $\Sigma$ and $B$, we derive the corresponding extrinsic curvatures, and in particular, for the latter, we have,  
\begin{align}\label{BY_GR_Eq09}
k&=-\frac{1}{\sqrt{h}}\partial _{\mu}\left(\sqrt{h}n^{\mu}\right)=-\partial _{r}n^{r}-n^{r}\partial _{r}\ln \sqrt{h}
\nonumber
\\
&=-\partial _{r}\sqrt{f}-\sqrt{f}\partial _{r}\ln \left(\frac{r^{2}\sin \theta}{\sqrt{f}}\right)
=-\frac{2}{r}\sqrt{f(r)}~.
\end{align}
The embedding of $\mathcal{B}$ in flat spacetime is trivial, and the trace of extrinsic curvature $k_{0}$ is $-2/r$. Using the expression for $k$ from \ref{BY_GR_Eq09} the Brown-York energy defined in \ref{BY_GR_Eq01} becomes
\begin{align}\label{BY_GR_Eq06}
E_{\rm{BY}}=\frac{1}{4\pi}\int d\theta d\phi ~r^{2}\sin \theta \frac{1}{r}\left(1-\sqrt{f(r)}\right)
=r\left(1-\sqrt{f(r)}\right)\nonumber\\=r\left(1-\sqrt{1- 2M/r}\right)~.
\end{align}
We expand it for large $r$ and write 
\begin{align}\label{BY_Asymp1}
E_{M}\approx M~.
\end{align}
Here, $E_M$ denotes the non-gravitational (matter) contribution, identified with the ADM mass parameter of the exterior solution. The gravitational field energy exterior to a sphere of radial extent $r$ can be derived as \cite{Dadhich:2016fku}
\begin{equation}
E_{\rm GF}=E_{\rm BY}-M=r\left(1-\sqrt{1- 2M/r}\right)-M~.
\end{equation}
We now impose the condition $E_{\rm GF} = k\, E_{M}$, where $k=1$ corresponds to a black hole and $k=1/2$ to a Buchdahl star. Varying $E_{\rm GF}$ with respect to the relevant parameters, we obtain
\begin{equation}
\delta E_{\rm GF}=\delta r\left(1-\sqrt{1- 2M/r}\right)+\bigg(\frac{-\delta M+\frac{M}{r}\delta r}{\sqrt{1- 2M/r}}\bigg)-\delta M~.
\end{equation}
The condition $E_{\rm GF}=k\,E_{M}$ further implies $M/r=2k/(k+1)^{2}$, which in turn gives the constraint $\delta M-(M/r)\,\delta r=0$, and as a consequene, we find that
\begin{equation}
\delta E_{\rm GF}=k \delta M~.
\end{equation}
Ths is indeed the equilibrium configuration characterizing the stability of the object, derived in a fully general sense, regardless of the nature of the perturbations. Next, we consider a scenario where the exterior metric is described by a Reissner-Nordstrom metric 
\begin{equation}
ds^{2}=-f(r) dt^{2}+\frac{d r^2}{f(r)}+ r^2(d \theta^{2}+\sin^2\theta)~.
\end{equation}
Here $f(r)$ is given by,
\begin{eqnarray}\label{charged}
    f &=& \bigg(1-\frac{2M}{r}+\frac{Q^{2}}{r^{2}}\bigg)~.
\end{eqnarray}
Once again, taking the hypersurface $\Sigma$ to be a $t=\textrm{constant}$ hypersurface and $B$ as a $r=\textrm{constant}$ surface within $\Sigma$, we derive the unit timelike normal $u_{a}$ to $\Sigma$ and unit spacelike normal $n_{a}$ to $B$ as 
\begin{align}
u_{a}&=-\sqrt{f}\left(1,0,0,0\right);\qquad u^{a}=\frac{1}{\sqrt{f}}\left(1,0,0,0\right)
\label{BY_GR_Eq07}
\\
n_{a}&=\frac{1}{\sqrt{f}}\left(0,1,0,0\right);\qquad n^{a}=\sqrt{f}\left(0,1,0,0\right)~.
\label{BY_GR_Eq08}
\end{align}
The corresponding extrinsic curvatures can be derived as
\begin{align}\label{BY_GR_Eq10}
k&=-\frac{1}{\sqrt{h}}\partial _{\mu}\left(\sqrt{h}n^{\mu}\right)=-\partial _{r}n^{r}-n^{r}\partial _{r}\ln \sqrt{h}
\nonumber
\\
&=-\partial _{r}\sqrt{f}-\sqrt{f}\partial _{r}\ln \left(\frac{r^{2}\sin \theta}{\sqrt{f}}\right)
=-\frac{2}{r}\sqrt{f(r)}~.
\end{align}
Using the expression for $k$ from \ref{BY_GR_Eq10}, the Brown-York energy becomes
\begin{align}\label{BY_GR_Eq11}
E_{\rm{BY}}=\frac{1}{4\pi}\int d\theta d\phi ~r^{2}\sin \theta \frac{1}{r}\left(1-\sqrt{f(r)}\right)
=r\left(1-\sqrt{f(r)}\right)\nonumber\\=r\left(1-\sqrt{1- 2M/r+Q^2/r^2}\right)~.
\end{align}
Once again, expanding it for large $r$ we write 
\begin{align}\label{BY_Asymp2}
E_{M}\approx M-\frac{Q^2}{2 r}~.
\end{align}
The gravitational field energy exterior to a sphere of radial extent $r$ can be derived as  \cite{Dadhich:2016fku},
\begin{equation}
E_{\rm GF}=E_{\rm BY}-(M-Q^2/2 r)=r\left(1-\sqrt{1- 2M/r+Q^2/r^2}\right)-(M-Q^2/2 r)~.
\end{equation}
Imposing the condition $E_{\rm GF} = k\,E_M$, where $k=1$ corresponds to a black hole and $k=\tfrac12$ to a Buchdahl star and taking the variation of $E_{\rm GF}$, we obtain
\begin{equation}
\delta E_{\rm GF} = \delta r \left( 1 - \sqrt{ 1 - \frac{2M}{r} + \frac{Q^2}{r^2} } \right) + \frac{-\delta M + \frac{M}{r} \delta r + \frac{Q\,\delta Q}{r} - \frac{Q^2}{r^2} \delta r}{\sqrt{ 1 - \frac{2M}{r} + \frac{Q^2}{r^2} }} - \delta \left( M - \frac{Q^2}{2r} \right).
\end{equation}
The relation $E_{\rm GF} = k E_M$ further implies
\begin{equation}
\frac{M - \frac{Q^2}{2r}}{r} = \frac{2k}{(k+1)^2},
\end{equation}
which leads to the constraint
\begin{equation}
\delta M - \frac{M}{r}\, \delta r - \frac{Q}{r}\, \delta Q + \frac{Q^2}{r^2}\, \delta r = 0.
\end{equation}
As a result, the variation of the gravitational field energy reduces to
\begin{equation}
\delta E_{\rm GF} = k\, \delta \left( M - \frac{Q^2}{2r} \right),
\end{equation}
i.e., $\delta E_{\rm GF}= k \delta E_M$. We therefore conclude once more, that the equilibrium configuration describing the object is stable in a fully general sense, independent of the specific nature of the perturbations. We can extend this analysis to pure Lovelock gravity, where the gravitational field energy is given by \cite{Chakraborty:2015kva, Gravanis:2010zz}
\begin{equation}
E_{\rm GF} = r^{\alpha} \left( 1 - \sqrt{ 1 - \frac{2M}{r^{\alpha}} } \right) - M,
\end{equation}
with $E_M = M$ and $\alpha = (d-2N-1)/N$, where $d$ is the spacetime dimension and $N$ is the Lovelock order. Taking the variation, we find
\begin{equation}
\delta E_{\rm GF} = \alpha r^{\alpha-1}\, \delta r \left( 1 - \sqrt{ 1 - \frac{2M}{r^{\alpha}} } \right) 
- r^{\alpha} \frac{-\delta M/r^{\alpha} + \alpha M \delta r/r^{\alpha+1}}{\sqrt{ 1 - \frac{2M}{r^{\alpha}} }} - \delta M~.
\end{equation}
Using $E_{\rm GF} = k E_M$ gives $M/r^{\alpha} = 2 k/(k+1)^2$, which implies
\begin{equation}
\delta M - \frac{\alpha M}{r} \delta r = 0~.
\end{equation}
Consequently, the variation of the gravitational field energy becomes
\begin{equation}
\delta E_{\rm GF} = k\, \delta M~,
\end{equation}
demonstrating the general stability of the equilibrium configuration in pure Lovelock gravity. Finally, for a rotating black hole, the gravitational field energy can be written as \cite{Chakraborty:2015kva}
\begin{equation}
E_{\rm GF} = r \left( 1 - \sqrt{ 1 - \frac{2M}{r} + \frac{a^2}{r^2} } \right) - \frac{M}{1 + a^2/r^2}~,
\end{equation}
where $E_M = M/(1 + a^2/r^2)$. Taking its variation one finds that
\begin{equation}
\delta E_{\rm GF} = \delta r \left( 1 - \sqrt{ 1 - \frac{2M}{r} + \frac{a^2}{r^2} } \right)
+ \frac{-\delta M + \frac{M}{r}\delta r + \frac{a\,\delta a}{r} - \frac{a^2}{r^2} \delta r}{\sqrt{ 1 - \frac{2M}{r} + \frac{a^2}{r^2} }}
- \delta \left( \frac{M}{1 + a^2/r^2} \right).
\end{equation}
The horizons of the rotating black hole are located at $r_\pm = M \pm \sqrt{M^2 - a^2}$, so that $M = (r_+ + r_-)/2$ and $a^2 = r_+ r_-$. At the horizon, the variation reduces to $\delta E_{\rm GF} = \delta r_+/2$. Similarly, the variation of the non-gravitational energy $\delta E_M$ also gives $\delta r_+/2$. Hence, on the horizon, we find that
\begin{equation}
\delta E_{\rm GF} = \delta E_M,
\end{equation}
demonstrating that the equilibrium of a rotating black hole is consistent with the general relation between gravitational and non-gravitational energies.

\section{Connection of Canonical Energy Extremization with Stability}\label{connection}
Thus far we have argued that the stability of objects attaining limiting compactness can be understood in a unified manner through the interplay between gravitational field energy and non-gravitational mass-energy. In the Brown-York framework, the gravitational field energy is defined as $E_{GF}(r, M) = E_{BY}(r, M) - E_M(r, M)$, where $E_M$ denotes the non-gravitational energy associated with the matter content. Black holes and Buchdahl stars are characterized by the compactness condition
\begin{equation}
E_{GF} = k\,E_M,
\label{eq:kcondition}
\end{equation}
with $k=1$ for black holes and $k=\tfrac12$ for Buchdahl stars. This relation encodes the energetic balance at the threshold of maximal compactness. To formalize this balance, we introduce the functional
\begin{equation}
\mathcal{F}(r,M) = E_{GF}(r,M) - k\,E_M(r,M),
\end{equation}
whose equilibrium configuration satisfies
\begin{equation}
\mathcal{F}(r_0,M_0) = 0.
\end{equation}
We now show that the limiting-compactness condition corresponds to an extremum of $\mathcal{F}$ and therefore signals stability in a global energetic sense. The first variation of $\mathcal{F}$ under arbitrary variations of $(r,M)$ is
\begin{equation}
\delta \mathcal{F} = \frac{\partial \mathcal{F}}{\partial r}\,\delta r + \frac{\partial \mathcal{F}}{\partial M}\,\delta M
= \delta E_{GF} - k\,\delta E_M.
\label{eq:deltaF1}
\end{equation}
Explicit evaluation for the Schwarzschild and Reissner-Nordstrom geometries shows that, in all cases,
\begin{equation}
\delta E_{GF} = k\,\delta E_M.
\label{eq:variationrelation}
\end{equation}
As a consequence,
\begin{equation}
\delta\mathcal{F}(r_0,M_0) = 0,
\end{equation}
demonstrating that the limiting-compactness configuration corresponds to a stationary point of the quasi-local energy functional. This result is independent of the matter equation of state, the specific form of the perturbation, or the presence of charge. Furthermore, direct evaluation of the second variation shows that the stationary point is a minimum, $\delta^2 \mathcal{F} > 0$, thereby establishing dynamical stability. At equilibrium, the first-variation condition induces a constraint relating the variations of $M$ and $r$. For the Schwarzschild case, one finds
\begin{equation}
\delta M = \frac{M_0}{r_0}\,\delta r,
\label{eq:constraintsch}
\end{equation}
while for the Reissner-Nordstrom geometry, the corresponding relation is
\begin{equation}
\delta M - \frac{M_0}{r_0}\delta r - \frac{Q_0}{r_0}\delta Q + \frac{Q_0^2}{r_0^2}\delta r = 0.
\label{eq:constraintrn}
\end{equation}
These relations show that admissible perturbations are constrained to lie along a one-dimensional curve in the $(r, M)$ configuration space. The second variation, evaluated along this constrained direction, gives
\begin{equation}
\delta^2 \mathcal{F} = \left(\frac{d^2 \mathcal{F}}{dr^2}\right)(\delta r)^2.
\end{equation}
As an example, for the Schwarzschild spacetime, the energy functional takes the explicit form
\begin{equation}
\mathcal{F}(r,M) = r\left(1-\sqrt{1-\frac{2M}{r}}\right) - (1+k)\,M.
\label{eq:Fsch}
\end{equation}
The equilibrium condition \ref{eq:kcondition} implies that
\begin{equation}
\frac{M_0}{r_0} = \frac{2k}{(1+k)^2},
\end{equation}
and, together with the constraint \ref{eq:constraintsch}, reduces the functional to an effective one-parameter function $\mathcal{F}(r)$. Using the equilibrium relation, we may write
\begin{equation}
\sqrt{1-\frac{2M_0}{r_0}} = \frac{|1-k|}{1+k}.
\end{equation}
Substituting this into \ref{eq:Fsch} yields
\begin{equation}
\mathcal{F}(r) = r\left(1-\frac{|1-k|}{1+k}\right) - \frac{2k}{1+k}\,r,
\end{equation}
from which one can derive
\begin{equation}
\frac{d^2\mathcal{F}}{dr^2} = \frac{2k}{r(1+k)^3}.
\end{equation}
Since $r > 0$ and $k > 0$ for all physically relevant configurations,
\begin{equation}
\frac{d^2\mathcal{F}}{dr^2} > 0,
\end{equation}
establishing that the equilibrium corresponds to a strict minimum of the energy functional. Considering a perturbed radius $r(t) = r_0 + \eta(t)$ and expanding $\mathcal{F}$ about $r_0$ we find that
\begin{equation}
\mathcal{F}(r) = \mathcal{F}(r_0) + \frac{1}{2}\left(\frac{d^2\mathcal{F}}{dr^2}\right)_{r_0}\eta^2 + \mathcal{O}(\eta^3).
\end{equation}
The effective potential governing radial perturbations is therefore
\begin{equation}
V_{\rm eff}(\eta) = \frac12 \left(\frac{d^2\mathcal{F}}{dr^2}\right)_{r_0}\eta^2,
\end{equation}
leading to the equation of motion
\begin{equation}
\ddot{\eta} + \omega^2 \eta = 0, \qquad \omega^2 = \left(\frac{d^2\mathcal{F}}{dr^2}\right)_{r_0} > 0.
\end{equation}
The perturbations therefore execute stable oscillations with real frequency, demonstrating that the extremum of $\mathcal{F}$ corresponds to a dynamically stable configuration. In this sense, objects saturating the limiting-compactness condition represents energetically selected and dynamically stable endpoints of a gravitational collapse.   \\

This extremization framework can also provide a natural starting point for extending the analysis beyond strictly static configurations. The stability of limiting compactness objects should persist under slow, controlled time evolution, provided the system remains close to the instantaneous extremum of the quasi-local energy functional $\mathcal{F}$. In the following section, we demonstrate this idea by considering a slowly evolving exterior geometry, for which each constant-time slice can be regarded as an instantaneous equilibrium configuration. We show that the condition $\delta\mathcal{F} = 0$ continues to encode the instantaneous limiting compactness relation, while the positivity of the second variation naturally leads to a time-dependent stability criterion. 

\section{Stability for a Non-Static Compact Object}
\label{sec:BY_conformal}
In a realistic gravitational collapse or quasi-equilibrium scenarios, the exterior spacetime does not remain exactly static but evolves slowly as the compact object adjusts towards (or away from) a limiting configuration. Our goal is to capture a departure from strict stationarity avoiding full dynamical complexity. To this end, we construct an evolving exterior geometry by performing a time-dependent conformal rescaling of a Schwarzschild seed metric as, 
\begin{equation}
g_{ab}=\Omega^2(t,r),\bar g_{ab},\qquad \bar g_{ab} dx^a dx^b = -f(r) dt^2 + \frac{dr^2}{f(r)} + r^2 d\Omega_{(2)}^2,
\label{eq:conf_metric_repeat}
\end{equation}
where $f(r) = 1 - 2M/r$ and $M$ denotes the mass parameter of the Schwarzschild seed. The conformal factor $\Omega(t,r)$ is assumed to vary slowly in time, encoding a quasi-static evolution of the exterior geometry. This allows the spacetime to evolve while remaining arbitrarily close to Schwarzschild on each constant-time slice. The areal radius of the physical spacetime is then given by $R(t,r) = \Omega(t,r)r$, which naturally incorporates both the slow temporal evolution and the radial deformation of the two-spheres. This construction provides a minimal and controlled extension of the static Schwarzschild exterior, sufficient to test the persistence of the limiting-compactness and stability conditions under slow dynamical evolution. The Brown-York quasi-local energy associated with a closed two-surface $B$ embedded in the spatial hypersurface $\Sigma:\,t=\mathrm{const}$ is, as discussed,
\begin{equation}
E_{BY}=\frac{1}{8\pi}\int_{B} d^2x\sqrt{q}\,(k-k_0),
\label{eq:BY_def_repeat}
\end{equation}
where $q_{AB}$ is the induced metric on $B$, $k$ is the trace of the extrinsic curvature of $B$ as embedded in $\Sigma$ and $k_0$ is the corresponding trace for an embedding of $B$ in flat space, serving as the reference \cite{Brown:1992br}. On a constant-time slice, the induced three-metric takes the form
\begin{equation}
h_{ij}dx^i dx^j = \frac{\Omega^2(t,r)}{f(r)}\,dr^2 + \Omega^2(t,r)\,r^2 d\Omega_{(2)}^2,
\end{equation}
with determinant $h = \frac{\Omega^6 r^4 \sin^2\theta}{f}$. The outward-pointing unit normal to the two-sphere $B$ within $\Sigma$ is purely radial and is given by
\begin{equation}
n^\mu = \Big(0,\; \frac{\sqrt{f}}{\Omega},\;0,\;0\Big),
\qquad
n_\mu = \Big(0,\;\frac{\Omega}{\sqrt{f}},\;0,\;0\Big).
\end{equation}
The trace of the extrinsic curvature is computed using the standard expression
\begin{equation}
k = -\frac{1}{\sqrt{h}}\,\partial_\mu\!\big(\sqrt{h}\,n^\mu\big) = -\,\frac{2\sqrt{f}}{\Omega^2 r}\,\Big(1 + \frac{r\Omega_{,r}}{\Omega}\Big).
\label{eq:k_conformal}
\end{equation}
For the reference term, we embed the two-sphere in flat space, using the areal radius $R = \Omega r$. The trace of the extrinsic curvature of a two-sphere of areal radius $R$ in flat space is then
\begin{equation}
k_0 = -\frac{2}{R} = -\frac{2}{\Omega r}.
\label{eq:k0}
\end{equation}
The area element of the two-surface $B$ induced from the spatial metric is
\begin{equation}
\sqrt{q}\,d^2x = \Omega^2 r^2\sin\theta\,d\theta\,d\phi.
\end{equation}
Substituting \ref{eq:k_conformal} and \ref{eq:k0} into \ref{eq:BY_def_repeat}, and performing the angular integration, we derive after simplification
\begin{equation}
E_{BY}(t,r) = \Omega r \Big[\,1 - \sqrt{f(r)}\,\Big(1 + \frac{r\Omega_{,r}(t,r)}{\Omega(t,r)}\Big)\Big] = R\Big[1 - \sqrt{f}\big(1 + r\partial_r\ln\Omega\big)\Big].
\label{eq:E_BY_conf}
\end{equation}
This expression provides the Brown-York quasi-local energy for a conformally rescaled Schwarzschild exterior evaluated on a constant-time slice. As a consistency check, one can check that the result reduces smoothly to the standard Schwarzschild Brown-York energy $E_{BY}=r(1-\sqrt{f})$, when the conformal factor $\Omega$ is independent of the radial coordinate. (We note here, that the derivation relies exclusively on spatial derivatives evaluated on a fixed $t$ slice. The time dependence of the conformal factor $\Omega(t,r)$ enters only parametrically through $t$. Under a quasi-static assumption, time-derivative contributions to the extrinsic geometry of the $t = \mathrm{const}$ hypersurfaces are subleading). As in the static treatment, we separate the quasi-local energy into non-gravitational and gravitational contributions. Denoting by $E_M(t,R)$ the non-gravitational energy contained within a two-sphere of areal radius $R$, we define the gravitational field energy exterior to radius $R$ as
\begin{equation}
E_{GF}(t,r) = E_{BY}(t,r) - E_M(t,R).
\end{equation}
This decomposition allows us to formulate a dynamical analogue of the limiting-compactness condition directly in terms of quasi-local quantities defined on each constant-time slice. We impose the instantaneous limiting-compactness condition
\begin{equation}
E_{GF}(t,r_0(t)) = k\,E_M(t,R_0(t)), \qquad 0<k\le1,
\label{eq:dyn_k_BY}
\end{equation}
which states that, at any given time, the gravitational field energy exterior to the boundary is proportional to the enclosed non-gravitational energy. This condition is equivalently expressed as the algebraic equilibrium relation
\begin{equation}
E_{BY}(t,r_0) = (1+k)\,E_M(t,R_0).
\label{eq:BY_equil} 
\end{equation}
Substituting \ref{eq:E_BY_conf} into \ref{eq:BY_equil}, we find the instantaneous algebraic condition determining the equilibrium coordinate radius $r_0(t)$, or equivalently, the areal radius $R_0(t)=\Omega r_0$,
\begin{equation}
R_0\Big[1 - \sqrt{f(r_0)}\big(1 + r_0\partial_r\ln\Omega|_{r_0}\big)\Big] = (1+k)\,E_M(t,R_0).
\label{eq:BY_algebraic}
\end{equation}
This relation represents the Brown-York generalization of the familiar static compactness condition $M/R = 2k/(1+k)^2$. To examine stability, we consider a small radial perturbation of the boundary at fixed time $t$,
\begin{equation}
r(t)=r_0(t)+\delta r, \qquad
R(t)=R_0(t)+\delta R, \qquad 
\delta R=\Omega(r_0)\,\delta r + r_0\Omega_{,r}\delta r,
\end{equation}
together with a corresponding perturbation $\delta E_M$ of the enclosed matter energy. We introduce the functional $\mathcal{F} = E_{GF} - kE_M$, whose vanishing characterizes the limiting-compactness configuration. The first variation of $\mathcal{F}$ on a constant $t$ slice is
\begin{equation}
\delta\mathcal{F}\big|_{t=\mathrm{const}} = \big(\partial_r E_{BY} - (1+k)\partial_r E_M\big)\,\delta r.
\label{eq:deltaF_BY}
\end{equation}
Using \ref{eq:E_BY_conf}, the radial derivative of the Brown-York energy at fixed $t$ is given by,
\begin{align}
\partial_r E_{BY}
&= \partial_r\!\big[\,\Omega r\big]\,
\Big(1-\sqrt{f}(1+r\partial_r\ln\Omega)\Big)
- \Omega r\,\partial_r\!\Big[\sqrt{f}(1+r\partial_r\ln\Omega)\Big] \nonumber\\
&= \Omega\Big(1+r\partial_r\ln\Omega\Big)
\Big[1-\sqrt{f}(1+r\partial_r\ln\Omega)\Big]
-\Omega r\,\partial_r\!\Big[\sqrt{f}(1+r\partial_r\ln\Omega)\Big].
\label{eq:dEby_dr}
\end{align}
After simplification, an instantaneous stationarity condition $\delta\mathcal{F} = 0$ leads us to write
\begin{equation}
\partial_r E_{BY}(t,r_0) = (1+k)\,\partial_r E_M(t,R_0),
\label{eq:BY_firstvar}
\end{equation}
which, together with \ref{eq:BY_algebraic}, constrains the admissible perturbations at time $t$. As in the static case, this relation may be rearranged to express the variation of the enclosed matter energy in terms of the radial displacement, or vice versa. To determine stability, we examine the second variation of $\mathcal{F}$ along this constrained direction. Defining the effective potential for radial motion as
\begin{equation}
V_{\rm eff}(t,r) := \mathcal{F}(t,r) = E_{BY}(t,r) - (1+k)E_M(t,R),
\end{equation}
we expand about the equilibrium radius $r_0(t)$ on a fixed $t$ slice,
\begin{equation}
V_{\rm eff}(t,r_0+\eta) = V_{\rm eff}(t,r_0) + \tfrac12 V''(t,r_0)\,\eta^2 + \mathcal{O}(\eta^3),
\end{equation}
where primes denote $\partial_r$ at fixed $t$. Linear stability requires
\begin{equation}
V''(t,r_0) > 0.
\label{eq:BY_stability}
\end{equation}
Differentiating \ref{eq:dEby_dr} and subtracting $(1+k)\partial_r^2 E_M$ yields
\begin{align}
V''(t,r_0)
&= \partial_r^2 E_{BY}(t,r_0)
- (1+k)\,\partial_r^2 E_M(t,R_0) \nonumber\\
&= \partial_r\Big\{
\Omega(1+r\partial_r\ln\Omega)
\big[1-\sqrt{f}(1+r\partial_r\ln\Omega)\big]
\Big\} - \partial_r\Big\{
\Omega r\,\partial_r\!\big[\sqrt{f}(1+r\partial_r\ln\Omega)\big]
\Big\} - (1+k)\partial_r^2 E_M,
\label{eq:Vpp_BY}
\end{align}
where $f'=2M/r^2$. For a quasi-static assumption
\begin{equation}
\dot\Omega(t,r)\equiv \partial_t\Omega = \mathcal{O}(\varepsilon), \qquad \varepsilon\ll1,
\end{equation}
we can consistently neglect terms of order $\varepsilon$ or higher in the instantaneous extrinsic geometry. Equivalently, the $t=\mathrm{const}$ slices can be treated as instantaneously time-symmetric to leading order. Under this approximation, the expressions (\ref{eq:E_BY_conf}-\ref{eq:Vpp_BY}) remain valid to leading order, with time entering only parametrically. For a more practical implementation of the stability analysis one can first solve the algebraic equilibrium condition \ref{eq:BY_algebraic} for $r_0(t)$, compute $V''(t,r_0)$ from \ref{eq:Vpp_BY} for a chosen conformal factor $\Omega(t,r)$ and then infer instantaneous stability or instability according to the sign of $V''(t,r_0)$.  \\

In the static limit, $\Omega=\Omega(r)$ with $\dot\Omega=0$, the analysis reduces to the purely spatial Brown-York treatment and reproduces the Schwarzschild result when $\Omega \equiv 1$. If the conformal factor varies slowly in space, $|r\partial_r\ln\Omega|\ll1$, one finds to first order $E_{BY}\simeq R(1-\sqrt{f}) - R\sqrt{f}\,r\partial_r\ln\Omega$. The matter sector plays a direct role through $\partial_r^2 E_M$. Finally, for faster evolution where $\varepsilon$ is not small, time-derivative and mixed space-time terms in the extrinsic geometry must be retained; these typically introduce additional negative contributions to $V''$, signalling kinetic destabilization and necessitating a full dynamical treatment.

\section{Conclusion}\label{con}

In this letter, we have shown that black holes and Buchdahl stars, the two objects of limiting compactness in general relativity, admit a unified energetic description in terms of Brown-York quasi-local energy. By expressing the gravitational field energy as the difference between quasi-local and non-gravitational mass-energy, we demonstrated that both configurations satisfy a universal balance condition $E_{GF} = kE_M$, with $k = 1$ and $k = 1/2$, respectively. This relation is independent of the microscopic properties of matter and depends only on the global energetic structure of the spacetime.   \\

The central result of this work is that the limiting compactness condition corresponds to an extremum: a minimum of a quasi-local energy functional. This provides a non-perturbative, equation-of-state-independent criterion for stability, applicable to static, charged, rotating, and higher-curvature (Lovelock) generalizations. Stability emerges here not as a consequence of specific perturbative dynamics, but as an energetic necessity dictated by the Einstein equations. We further outline how this canonical energy extremization principle naturally extends to slowly evolving configurations. In the quasi-static regime, each constant-time slice of an evolving exterior geometry may be regarded as an instantaneous equilibrium configuration, with stability controlled by the same energy extrema. This perspective establishes a direct conceptual link between static stability and slow gravitational collapse and clarifies why objects that saturate the limiting compactness represent dynamically selected and energetically stable endpoints of gravitational compression.  \\

Our results highlight the utility of quasi-local gravitational energy as a fundamental diagnostic for compactness, equilibrium, and stability in relativistic gravitating systems, and suggest that energetic extremization may provide a unifying principle for understanding stability across a broad class of gravitational theories.

\section*{Acknowledgement}
SC acknowledges the IUCAA for providing the facility and support under the visiting associateship program. Acknowledgement is given to the Vellore Institute of Technology for the financial support through its Seed Grant (No. SG20230027), 2023.

\bibliography{references}

\bibliographystyle{apsrev4-1}

\end{document}